\documentclass[showpacs,aps,graphicx,twocolumn]{revtex4}
\usepackage{graphicx}

\begin{document}
\title{Quantum state sharing of an arbitrary two-qubit state
with two-photon entanglements and Bell-state measurements}
\author{ Fu-Guo Deng,$^{1,2,3}$\footnote{ Email address: fgdeng@bnu.edu.cn}
Xi-Han Li,$^{1,2}$ Chun-Yan Li,$^{1,2}$
Ping Zhou,$^{1,2}$ and Hong-Yu Zhou$^{1,2,3}$ }
\address{$^1$ Key Laboratory of Beam Technology and Material
Modification of Ministry of Education, Beijing Normal University,
Beijing 100875,
People's Republic of China\\
$^2$ Institute of Low Energy Nuclear Physics, and Department of
Material Science and Engineering, Beijing Normal University,
Beijing 100875,
People's Republic of China\\
$^3$ Beijing Radiation Center, Beijing 100875,  People's Republic
of China}
\date{\today }

\begin{abstract}
Two schemes for sharing an arbitrary two-qubit state based on
entanglement swapping are proposed with Bell-state measurements
and local unitary operations. One is based on the quantum channel
with four Einstein-Podolsky-Rosen (EPR) pairs shared in advance.
The other is based on a circular topological structure, i.e., each
user shares an EPR pair with his neighboring one. The advantage of
the former is that the construction of the quantum channel between
the agents is controlled by the sender Alice, which will improve
the security of the scheme. The circular scheme reduces the
quantum resource largely when the number of the agents is large.
Both of those schemes have the property of high efficiency as
almost all the instances can be used to split the quantum
information. They are more convenient in application than the
other schemes existing as they require only two-qubit
entanglements and two-qubit joint measurements for sharing an
arbitrary two-qubit state.

\end{abstract}

\maketitle

\section{introduction}
\label{section1}
The basic idea of secret sharing \cite{Blakley}
in a simple case is that a secret ($M_A$) is divided by the sender
Alice into two pieces which will be distributed to two parties,
Bob and Charlie, respectively, and they can reconstruct the secret
if and only if both act in concert. Each can get nothing about the
message $M_A$. As classical signal can be copied freely and fully
without leaving a track, there is no way for people to complete
the task unconditionally securely with classical physics in
principle. When quantum mechanics enters the field of information,
the case is changed. Quantum secret sharing (QSS), an important
branch of quantum communication, is the generalization of
classical secret sharing into quantum scenario \cite{HBB99,KKI}.
There are three main goals in QSS. The first one is used to
distribute a private key among several parties, such as those in
Refs. \cite{HBB99,KKI,guoqss,longqss,YanGao,CPY,CabelloAS}; the
second is used for splitting a classical secret
\cite{HBB99,KKI,Gottesman,Bandyopadhyay,Karimipour,zhangswapping,zhanglm,zhangPLA,cleve},
and the third one can be used to share an unknown quantum state
\cite{Peng,dengmQSTS,dengpra}, which has to resort to quantum
entanglement.

Certainly, quantum key distribution (QKD) provides a secure way for
generating a private key between two remote parties and then can be
used to complete the task for disturbing the key among several
parties. The difference between QSS and QKD \cite{Gisin} is that the
former can reduce the resource necessary to implement multi-party
secret sharing tasks and is more convenient than that with QKD
\cite{guoqss}. A pioneering QSS scheme was proposed by Hillery,
Bu\v{z}ek and Berthiaume in 1999 by using three-particle and
four-particle entangled Greenberger-Horne-Zeilinger (GHZ) states for
sharing classical information, called HBB99 customarily for short.
For sharing a quantum secret, almost all of the existing QSS
protocols are either used to split a single qubit \cite{Peng} or
resort to $m$-particle entanglements \cite{dengpra} and $m$-particle
joint measurements ($m>2$) \cite{Peng,dengmQSTS}. The producing and
measurement of $m$-particle entanglement both are not easy with
present techniques
\cite{Mentanglement1,Mentanglement2,Mentanglement3}.

From the view of security, sharing a quantum secret in QSS is
similar to quantum secure direct communication (QSDC)
\cite{twostep,QOTP,Wangc} in which the secret message is
transmitted directly without creating a private key and then
encrypting the message as the quantum secret should not be leaked
to the dishonest one. It is necessary for QSS to set up a quantum
channel securely in advance
\cite{HBB99,KKI,Bandyopadhyay,Karimipour,zhangPLA,Peng,cleve},
which is same as QSDC in Refs. \cite{twostep,QOTP,Wangc}. The way
for sharing a sequence of two-particle maximally entangled states,
Einstein-Podolsky-Rosen (EPR) pairs is discussed in Refs
\cite{twostep,guoatom}.

In this paper, we will present a quantum state sharing (which is
abbreviated as \emph{QSTS} in Ref.\cite{dengmQSTS}, different from
QSS for classical information)
scheme for sharing an arbitrary two-qubit state $\vert \chi\rangle_{ab}=\alpha \vert%
00\rangle_{ab} + \beta \vert 01\rangle_{ab} + \gamma \vert
10\rangle_{ab} + \delta \vert 11\rangle_{ab}$  based on
entanglement swapping \cite{swapping1,swapping2,swapping3} with
Bell-state measurements and local unitary operations. It will be
shown that the state $\vert \chi\rangle_{ab}$ can be split by two
agents with  four EPR pairs shared in advance and four Bell-state
measurements, not $m$-particle joint measurements ($m>2$). Any one
in the two agents has the choice to reconstruct the original state
$\vert \chi\rangle_{ab}$ with the help of the other's. Moreover,
we present a circular topological structure for splitting the
state $\vert \chi\rangle_{ab}$ with EPR pairs and Bell-state
measurements efficiently as it reduces the quantum resource
largely when the number of the agents is large. Almost all the EPR
pairs can be used for quantum communication in those two schemes,
their efficiency for qubits approaches the maximal value, same as
Refs.\cite{Peng,dengmQSTS,dengpra}. They are more convenient in
application than the other schemes existing as they require only
two-qubit entanglements and two-qubit joint measurements, not GHZ
states, for sharing an arbitrary two-qubit state.

\section{QSTS protocol with EPR pairs and Bell-basis measurements}
\label{section2}
An EPR pair is in one of the four Bell states
shown as follows:
\begin{eqnarray}
\vert \psi ^{\pm}\rangle_{AB} =\frac{1}{\sqrt{2}}(\vert 0\rangle
_{A}\vert 1\rangle _{B}\pm\vert
1\rangle _{A}\vert 0\rangle _{B}), \label{EPR12}\\
\vert \phi ^{\pm}\rangle_{AB} =\frac{1}{\sqrt{2}}(\vert 0\rangle
_{A}\vert 0\rangle _{B}\pm\vert 1\rangle _{A}\vert 1\rangle _{B}).
\label{EPR34}
\end{eqnarray}
where $\vert 0\rangle$ and $\vert 1\rangle$ are the two
eigenvectors of two-level quantum system, such as the
polarizations of photon along the z-direction, say $\sigma_z$ .
The four local unitary operations $U_{i}$ ($i=0,1,2,3$) can
transform each one of the four Bell states into another.
\begin{eqnarray}
U_{0}=\vert 0\rangle \langle 0\vert +\vert 1\rangle \langle
1\vert, \,\,\,\,\, U_{1}=\vert 0\rangle \langle 0\vert
-\vert 1\rangle \langle 1\vert, \nonumber\\
U_{2}=\vert 1\rangle \langle 0\vert +\vert 0\rangle \langle
1\vert, \,\,\,\,\,\, U_{3}=\vert 0\rangle \langle 1\vert -\vert
1\rangle \langle 0\vert. \label{U}
\end{eqnarray}
For example,
\begin{eqnarray}
I \otimes U _{0}\vert \psi^-\rangle&=&\vert \psi^-\rangle,
\,\,\,\,\,\,\,\, I \otimes U_{1}\vert \psi^-\rangle=-\vert
\psi^+\rangle,\nonumber\\
I \otimes U _{2}\vert \psi^-\rangle&=&\vert \phi^-\rangle,
\,\,\,\,\,\,\,\, I \otimes U _{3}\vert \psi^-\rangle=\vert
\phi^+\rangle ,\label{Uexample}
\end{eqnarray}
where $I=U_0$ is the 2$\times$2 identity operator which means
doing nothing on the particle.

The basic idea of this QSTS scheme for splitting an entangled
state $\vert \chi\rangle_{ab}=\alpha \vert 00\rangle_{ab} + \beta
\vert 01\rangle_{ab} + \gamma \vert 10\rangle_{ab} + \delta \vert
11\rangle_{ab}$ based on entanglement swapping is shown in Fig.1.
Alice shares two EPR photon pairs $\vert \psi^-\rangle_{12}$ and
$\vert \psi^-\rangle_{34}$ with Bob, and another two pairs
$\vert\psi^-\rangle_{56}$ and $\vert \psi^-\rangle_{78}$ with
Charlie. She retains the photons $1$, $3$, $5$, $7$, and the two
photons $a$ and $b$ in the entangled state $\vert
\chi\rangle_{ab}$. Bob and Charlie keep the photons $2$ and $4$,
and $6$ and $8$, respectively. The joint state of the quantum
system composed of the six photons $a$, $b$, $3$, $4$, $5$, and
$6$ can be written as
\begin{widetext}
\begin{eqnarray}
\vert \Phi\rangle_{ab3456}&\equiv&(\alpha \vert 00\rangle + \beta
\vert 01\rangle + \gamma \vert 10\rangle + \delta \vert
11\rangle)_{ab}\otimes\vert \psi^-\rangle_{34}\otimes\vert
\psi^-\rangle_{56}\nonumber\\
&=&\frac{1}{4}\{\vert\psi^-\rangle_{a3}[\vert\psi^-\rangle_{b5}(\alpha
\vert 00\rangle + \beta \vert 01\rangle + \gamma \vert 10\rangle +
\delta \vert 11\rangle)_{46} + \vert \psi^+\rangle_{b5}(\alpha
\vert 00\rangle - \beta \vert 01\rangle + \gamma \vert 10\rangle -
\delta \vert 11\rangle)_{46}\nonumber\\
&&\,\,\,\,\,\,\,\,\,\,\,\,\,\,\,\,\,\,\,\,\,\,\, -
\vert\phi^-\rangle_{b5}(\alpha \vert 01\rangle + \beta \vert
00\rangle + \gamma \vert 11\rangle + \delta \vert 10\rangle)_{46}
- \vert \phi^+\rangle_{b5}(\alpha \vert 01\rangle - \beta \vert
00\rangle + \gamma \vert 11\rangle - \delta \vert
10\rangle)_{46}]\nonumber\\ 
&& \,\,\,\, +
\vert\psi^+\rangle_{a3}[\vert\psi^-\rangle_{b5}(\alpha \vert
00\rangle + \beta \vert 01\rangle - \gamma \vert 10\rangle -
\delta \vert 11\rangle)_{46} + \vert \psi^+\rangle_{b5}(\alpha
\vert 00\rangle - \beta \vert 01\rangle - \gamma \vert 10\rangle +
\delta \vert 11\rangle)_{46}\nonumber\\
&&\,\,\,\,\,\,\,\,\,\,\,\,\,\,\,\,\,\,\,\,\,\,\, -
\vert\phi^-\rangle_{b5}(\alpha \vert 01\rangle + \beta \vert
00\rangle - \gamma \vert 11\rangle - \delta \vert 10\rangle)_{46}
- \vert \phi^+\rangle_{b5}(\alpha \vert 01\rangle - \beta \vert
00\rangle - \gamma \vert 11\rangle + \delta \vert
10\rangle)_{46}]\nonumber\\ 
&& \,\,\,\, -
\vert\phi^-\rangle_{a3}[\vert\psi^-\rangle_{b5}(\alpha \vert
10\rangle + \beta \vert 11\rangle + \gamma \vert 00\rangle +
\delta \vert 01\rangle)_{46} + \vert \psi^+\rangle_{b5}(\alpha
\vert 10\rangle - \beta \vert 11\rangle + \gamma \vert 00\rangle -
\delta \vert 01\rangle)_{46}\nonumber\\
&&\,\,\,\,\,\,\,\,\,\,\,\,\,\,\,\,\,\,\,\,\,\,\, -
\vert\phi^-\rangle_{b5}(\alpha \vert 11\rangle + \beta \vert
10\rangle + \gamma \vert 01\rangle + \delta \vert 00\rangle)_{46}
- \vert \phi^+\rangle_{b5}(\alpha \vert 11\rangle - \beta \vert
10\rangle + \gamma \vert 01\rangle - \delta \vert
00\rangle)_{46}]\nonumber\\ 
&& \,\,\,\, -
\vert\phi^-\rangle_{a3}[\vert\psi^-\rangle_{b5}(\alpha \vert
10\rangle + \beta \vert 11\rangle - \gamma \vert 00\rangle -
\delta \vert 01\rangle)_{46} + \vert \psi^+\rangle_{b5}(\alpha
\vert 10\rangle - \beta \vert 11\rangle - \gamma \vert 00\rangle +
\delta \vert 01\rangle)_{46}\nonumber\\
&&\,\,\,\,\,\,\,\,\,\,\,\,\,\,\,\,\,\,\,\,\,\,\, -
\vert\phi^-\rangle_{b5}(\alpha \vert 11\rangle + \beta \vert
10\rangle - \gamma \vert 01\rangle - \delta \vert 00\rangle)_{46}
- \vert \phi^+\rangle_{b5}(\alpha \vert 11\rangle - \beta \vert
10\rangle - \gamma \vert 01\rangle + \delta \vert
00\rangle)_{46}]\}.
\end{eqnarray} \label{result5}
\end{widetext}
For splitting the state $\vert \chi\rangle_{ab}$, Alice first
performs Bell-state measurement on the photons $a$ and $3$, and
then $b$ and $5$. She records the results $R_{a3}$ and $R_{b5}$.
In this way, the state $\vert \chi\rangle_{ab}$ is transferred to
the particles 4 and 6 which are kept by Bob and Charlie,
respectively. In order to set up a quantum channel for Bob and
Charlie, Alice performs Bell-state measurement on the photons $1$
and $7$, and records the result $R_{17}$. With $R_{17}$, the state
of the photons $2$ and $8$ can be determined as
\begin{eqnarray}
\vert \Phi\rangle_{1278}&\equiv&\vert
\psi^-\rangle_{12}\otimes\vert
\psi^-\rangle_{78}\nonumber\\
&=&\frac{1}{2}(\vert\psi^-\rangle_{17}\vert\psi^-\rangle_{28} -
\vert \psi^+\rangle_{17}\vert
\psi^+\rangle_{28}\nonumber\\
&-& \vert\phi^-\rangle_{17}\vert\phi^-\rangle_{28} + \vert
\phi^+\rangle_{17}\vert \phi^+\rangle_{28}).
\end{eqnarray}

\begin{figure}[!h]
\begin{center}
\includegraphics[width=8cm,angle=0]{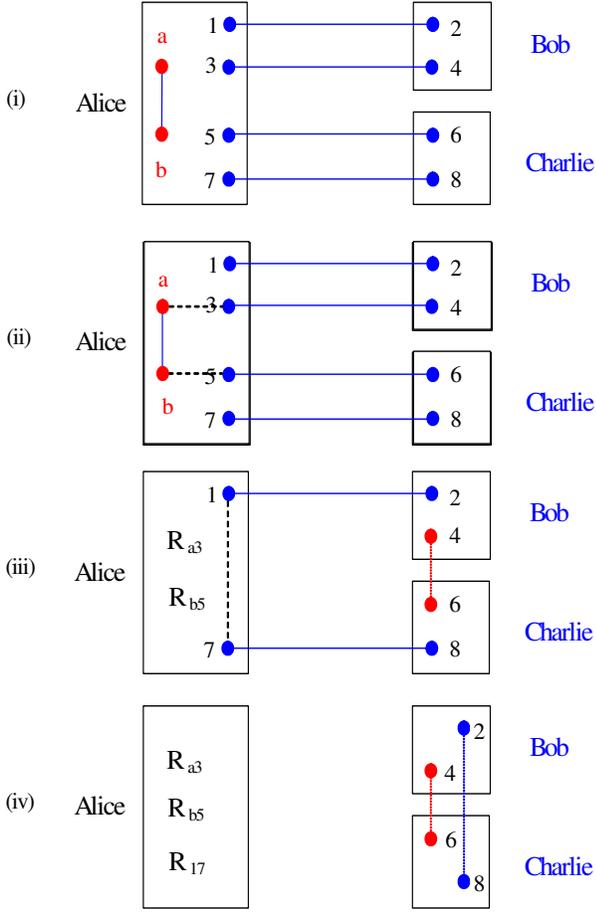} \label{f1}
\caption{ Quantum secret sharing based on entanglement swapping
with Bell-basis measurements and local unitary operation by using
four EPR pairs as the quantum channel. The bold lines connect
qubits in Bell states or the two-particle entangled state
$\vert\chi\rangle_{ab}$, the dashed lines connect qubits on which
a Bell measurement is made, and the diamond lines connect qubits
in entangled states (or Bell state) induced by entanglement
swapping, similar to that in Ref. \cite{cabello}. $R_{a3}$,
$R_{b5}$ and $R_{17}$ are the results of the Bell-basis
measurements on the particles a and 3, b and 5, 1 and 7,
respectively. }
\end{center}
\end{figure}

For reconstructing the original state $\vert \chi\rangle_{ab}$,
Bob or Charlie performs Bell-state measurement on his two photons
and then tells the other one the result when they act in concert.
We assume that Charlie will obtain the quantum secret message
$\vert \chi\rangle_{ab}$ with the help of Bob's. Due to symmetry,
the other cases are the same as it with or without a little of
modification. As an example, let us suppose that the results
$R_{a3}$, $R_{b5}$ and $R_{17}$ published by Alice are $\vert
\psi^-\rangle_{a3}$, $\vert \psi^-\rangle_{b5}$ and $\vert
\psi^-\rangle_{17}$.
\begin{eqnarray}
\vert \Phi\rangle_{2846}&\equiv&\vert
\psi^-\rangle_{28}\otimes(\alpha \vert 00\rangle + \beta \vert
01\rangle + \gamma \vert 10\rangle +
\delta \vert 11\rangle)_{46}\nonumber\\
&=&\frac{1}{2}\{\vert\psi^-\rangle_{24}(\alpha \vert 00\rangle +
\beta \vert 01\rangle + \gamma \vert 10\rangle + \delta \vert
11\rangle)_{86}\nonumber\\
&& \,\,-\vert\psi^+\rangle_{24}(\alpha \vert 00\rangle + \beta
\vert 01\rangle - \gamma \vert 10\rangle - \delta \vert
11\rangle)_{86}\nonumber\\
&& \,\, +\vert\phi^-\rangle_{24}(\alpha \vert 10\rangle + \beta
\vert 11\rangle + \gamma \vert 00\rangle + \delta \vert
01\rangle)_{86}\nonumber\\
&& \,\, +\vert\phi^+\rangle_{24}(\alpha \vert 10\rangle + \beta
\vert 11\rangle - \gamma \vert 00\rangle - \delta \vert
01\rangle)_{86}\}.\nonumber\\
\end{eqnarray}
If the result of the Bell-state measurement $R_{24}$ done by Bob
is $\vert\psi^-\rangle_{24}$, $\vert\psi^+\rangle_{24}$,
$\vert\phi^-\rangle_{24}$ or $\vert\phi^+\rangle_{24}$, Charlie
needs to perform the local unitary operations $U_0\otimes U_0$,
$U_1\otimes U_0$, $U_2\otimes U_0$ or $U_3\otimes U_0$ on the
particles 8 and 6 respectively,  and then reconstructs the state
$\vert\chi\rangle_{ab}$.

For the other cases, the relation between the local unitary
operations with which Bob can recover the original state
$\vert\chi\rangle_{ab}$ and the results $R_{a3}$, $R_{b5}$,
$R_{17}$ and $R_{24}$ is shown in Table I. Same as those in Ref.
\cite{dengpra}, we define $V$ as the bit value of the Bell state,
i.e., $V_{\vert \phi^\pm\rangle}\equiv 0$, $V_{\vert
\psi^\pm\rangle}\equiv 1$; That is, the bit value $V=0$ if the
states of the two particles in a Bell state are parallel,
otherwise $V=1$. $V_{total} \equiv V_{a3}\oplus V_{b5} \oplus
V_{17}\oplus V_{24}$. $P$ denotes the parity of the result of the
Bell-state measurement on the two-particle quantum system $R_i \in
\{\vert \psi^+\rangle, \vert \psi^-\rangle, \vert \phi^+\rangle,
\vert \phi^-\rangle \}$, i.e., $P_{\vert \psi
^\pm\rangle}\equiv\pm$, $P_{\vert \phi ^\pm\rangle}\equiv\pm$ and
$P_{total} \equiv \prod_{i=1} \otimes P_{R_i}=P_{R_{a3}}\otimes
P_{R_{b5}}\otimes P_{R_{17}}\otimes P_{R_{24}}$;  $\Phi_{86}$ is
the state of the particles $8$ and $6$ after all the Bell-basis
measurements are taken, $\oplus$ means summing modulo 2 and the
unitary operations $U_i\otimes U_j$ $(i,j \in \{0,1,2,3\})$
represents performing the unitary operation $U_i$ on the particle
$8$ and the operation $U_j$ on the particle $6$, respectively. For
instance, if the results of $R_{a3}$, $R_{b5}$, $R_{17}$ and
$R_{24}$ are $\vert \psi^-\rangle_{a3}$, $\vert
\phi^-\rangle_{b5}$, $\vert \psi^+\rangle_{17}$ and $\vert
\psi^-\rangle_{24}$, then $V_{total}=1\oplus 0\oplus 1\oplus 1=1$,
$V_{b5}=0$, $P_{b5}=-$, and $P_{total}=(-)\otimes (-) \otimes (+)
\otimes (-)=-$, and Charlie performs the unitary operations $U_1$
and $U_2$ on the particles $8$ and $6$ respectively for
reconstructing the original state $\vert\chi\rangle_{ab}$.

\begin{center}
\begin{table}[!h]
\caption{The relation between the local unitary operations and the
results $R_{a3}$, $R_{b5}$, $R_{17}$ and $R_{24}$. $\Phi_{86}$ is
the state of the two particles hold in the hand of Charlie after
all the measurements are done by Alice and Bob; $U_i\otimes U_j$
are the local unitary operations with which Charlie can
reconstruct the unknown state $\vert\chi\rangle_{ab}$. }
\begin{tabular}{ccccccc}\hline\hline
$V_{total}$  &  $V_{b5}$ &  $P_{b5}$ &  $P_{total}$ & $\Phi_{86}$
&  $U_i\otimes U_j$
\\\hline 
 0  & 1 &  $-$ &  $+$ &   $\alpha\vert
00\rangle + \beta\vert 01\rangle + \gamma\vert 10\rangle + \delta
\vert 11\rangle$ &  $U_0\otimes U_0 $
\\ 
 0  &  1 &  $+$ &  $-$ &
$\alpha\vert 00\rangle - \beta\vert 01\rangle + \gamma \vert
10\rangle - \delta \vert 11\rangle$ &  $U_0\otimes U_1 $
\\ 
1  & 0 & $-$ & $+$ &   $\alpha\vert 00\rangle + \beta\vert
01\rangle - \gamma\vert 10\rangle - \delta \vert 11\rangle$ &
$U_1\otimes U_0 $
\\ 
 1  & 0 & $+$ & $-$ &   $\alpha\vert
00\rangle - \beta\vert 01\rangle - \gamma\vert 10\rangle + \delta
\vert 11\rangle$ &  $U_1\otimes U_1 $
\\ 
 0  & 1 & $-$ & $-$ &   $\alpha\vert
01\rangle + \beta\vert 00\rangle + \gamma  \vert 11\rangle +
\delta\vert 10\rangle$ &  $U_0\otimes U_2 $
\\ 
 0  & 1 & $+$ & $+$ &  $\alpha\vert
01\rangle - \beta\vert 00\rangle + \gamma\vert 11\rangle - \delta
\vert 10\rangle$ &  $U_0\otimes U_3 $
\\ 
 1  & 0 & $-$ & $-$ &  $\alpha\vert
01\rangle + \beta\vert 00\rangle - \gamma\vert 11\rangle - \delta
\vert 10\rangle$ &  $U_1\otimes U_2 $
\\ 
 1  & 0 & $+$ & $+$ &  $\alpha\vert
01\rangle - \beta\vert 00\rangle - \gamma\vert 11\rangle + \delta
\vert 10\rangle$ &  $U_1\otimes U_3 $
\\ 
 1  & 1 & $-$ & $+$ &  $\alpha\vert
10\rangle + \beta\vert 11\rangle + \gamma\vert 00\rangle + \delta
\vert 01\rangle$ &  $U_2\otimes U_0 $
\\ 
 1  & 1 & $+$ & $-$ &  $\alpha\vert
10\rangle - \beta\vert 11\rangle + \gamma \vert 00\rangle - \delta
\vert 01\rangle$ &  $U_2\otimes U_1 $
\\ 
 0  & 0 & $-$ & $+$ &  $\alpha\vert
10\rangle + \beta\vert 11\rangle - \gamma\vert 00\rangle - \delta
\vert 01\rangle$ &  $U_3\otimes U_0$
\\ 
 0  & 0 & $+$ & $-$ &   $\alpha\vert
10\rangle - \beta\vert 11\rangle - \gamma\vert 00\rangle + \delta
\vert 01\rangle$ &  $U_3\otimes U_1 $
\\ 
 1  & 1 & $-$ & $-$ &  $\alpha\vert
11\rangle + \beta\vert 10\rangle + \gamma\vert 01\rangle + \delta
\vert 00\rangle$ &  $U_2\otimes U_2 $
\\ 
 1  & 1 & $+$ & $+$ &   $\alpha\vert
11\rangle - \beta\vert 10\rangle + \gamma\vert 01\rangle - \delta
\vert 00\rangle$ &  $U_2\otimes U_3 $
\\ 
 0  & 0 & $-$ & $-$ &   $\alpha\vert
11\rangle + \beta\vert 10\rangle - \gamma\vert 01\rangle - \delta
\vert 00\rangle$ &  $U_3\otimes U_2 $
\\ 
0  & 0 & $+$ & $+$ &   $\alpha\vert 11\rangle - \beta\vert
10\rangle -\gamma\vert 01\rangle + \delta \vert 00\rangle$ &
$U_3\otimes U_3 $
\\\hline \hline
\end{tabular}\label{table1}
\end{table}
\end{center}

In detail, Alice performs Bell-state measurements on the particles
$a$ and $3$, $b$ and $5$, $1$ and $7$, and she publishes the
results $R_{b5}$ , $R_{a3}\oplus R_{b5}\oplus R_{17}$ with simple
coding, i.e., 0 or 1, and the parities $P_{b5}$ and
$P=P_{a3}\otimes P_{b5}\otimes P_{17}$ ($+$ or $-$). She only pays
four bits of classical information for announcing her results in
public, not six bits. Subsequently, Bob takes Bell-state
measurement on the particles $2$ and $4$, and records the result
$R_{24}$ including its bit value and its parity (two bits of
classical information). With the four bits of information
published by Alice, Charlie can reconstruct the original state
$\vert \chi\rangle_{ab}$ according to the Table I with the help of
Bob's. On the other hand, neither Bob nor Charlie can obtain the
unknown two-qubit state if they do not cooperate even they get the
information published by Alice. Let us suppose that the result of
the measurement on particles $b$ and $5$ done by Alice is $\vert
\phi^+\rangle_{b5}$. From Table \ref{table11}, we can see that
Charlie has only the probability $\frac{1}{4}$ to choose two
correct local unitary operations for reconstructing the two-qubit
unknown state if he knows the information published by Alice after
Bob performed the Bell-state measurement on his two particles.
That is, the four results of Bob's measurements represent four
kinds of combination of the two operations on the two particles
kept by Charlie. Moreover, if Bob does not measures his two
particles, Charlie can only obtain a random result, no useful
information about the unknown state,  as he gets only a part of
the two-qubit quantum system $\vert \chi\rangle_{ab}$ after the
entanglement swapping is performed by Alice.

\begin{widetext}
\begin{center}
\begin{table}[!h]
\caption{The relation between the local unitary operations and the
results of the measurements done by Bob $R_{Bob}$ after Alice
published her information about her measurements. Here
$V_{Alice}=V_{a3} \oplus V_{b5} \oplus V_{17}$, $P_{Alice}=P_{a3}
\otimes P_{b5} \otimes P_{17}$, and  $U_i\otimes U_j$ are the
local unitary operations with which Charlie can reconstruct the
unknown state $\vert\chi\rangle_{ab}$.}
\begin{tabular}{cccccccc}\hline\hline
$V_{Alice}$  &  $V_{b5}$ &  $P_{b5}$ &  $P_{Alice}$ & $R_{Bob}$ &
$\Phi_{86}$ & $U_i\otimes U_j$
\\ \hline
0  & 0 & $+$ & $+$ & $\phi^+$ & $\alpha\vert 11\rangle -
\beta\vert 10\rangle -\gamma\vert 01\rangle + \delta \vert
00\rangle$ & $U_3\otimes U_3 $\\
 0  & 0 & $+$ & $+$ &  $\phi^-$ &  $\alpha\vert
10\rangle - \beta\vert 11\rangle - \gamma\vert 00\rangle + \delta
\vert 01\rangle$ &  $U_3\otimes U_1 $\\
 0  & 0 & $+$ & $+$ & $\psi^+$ & $\alpha\vert
01\rangle - \beta\vert 00\rangle - \gamma\vert 11\rangle + \delta
\vert 10\rangle$ &  $U_1\otimes U_3 $
\\
 0  & 0 & $+$ & $+$ & $\psi^-$ & $\alpha\vert
00\rangle - \beta\vert 01\rangle - \gamma\vert 10\rangle + \delta
\vert 11\rangle$ &  $U_1\otimes U_1 $
\\
 0  & 0 & $+$ & $-$ & $\phi^+$ &$\alpha\vert
10\rangle + \beta\vert 11\rangle - \gamma\vert 00\rangle - \delta
\vert 01\rangle$ &  $U_3\otimes U_0$
\\
 0  & 0 & $+$ & $-$ & $\phi^-$ & $\alpha\vert
11\rangle + \beta\vert 10\rangle - \gamma\vert 01\rangle - \delta
\vert 00\rangle$ &  $U_3\otimes U_2 $
\\
0  & 0 & $+$ & $-$ & $\psi^+$  &$\alpha\vert 00\rangle +
\beta\vert 01\rangle - \gamma\vert 10\rangle - \delta \vert
11\rangle$ & $U_1\otimes U_0 $
\\
0  & 0 & $+$ & $-$ & $\psi^-$ &$\alpha\vert 01\rangle + \beta\vert
00\rangle - \gamma\vert 11\rangle - \delta \vert 10\rangle$ &
$U_1\otimes U_2 $
\\
 1  & 0 & $+$ & $+$ & $\phi^+$ &$\alpha\vert
01\rangle - \beta\vert 00\rangle + \gamma\vert 11\rangle - \delta
\vert 10\rangle$ &  $U_0\otimes U_3 $
\\
 1  &  0 &  $+$ &  $+$ & $\phi^-$ &
$\alpha\vert 00\rangle - \beta\vert 01\rangle + \gamma \vert
10\rangle - \delta \vert 11\rangle$ &  $U_0\otimes U_1 $
\\
1  & 0 & $+$ & $+$ & $\psi^-$ &$\alpha\vert 10\rangle - \beta\vert
11\rangle + \gamma \vert 00\rangle - \delta \vert 01\rangle$ &
$U_2\otimes U_1 $
\\
1  & 0 & $+$ & $+$ & $\psi^+$  &$\alpha\vert 11\rangle -
\beta\vert 10\rangle + \gamma\vert 01\rangle - \delta \vert
00\rangle$ & $U_2\otimes U_3 $
\\
 1  & 0 &  $+$ &  $-$ &  $\phi^+$ &$\alpha\vert
00\rangle + \beta\vert 01\rangle + \gamma\vert 10\rangle + \delta
\vert 11\rangle$ &  $U_0\otimes U_0 $
\\
 1  & 0 & $+$ & $-$ &  $\phi^-$ &$\alpha\vert
01\rangle + \beta\vert 00\rangle + \gamma  \vert 11\rangle +
\delta\vert 10\rangle$ &  $U_0\otimes U_2 $
\\
 1  & 0 & $+$ & $-$ & $\psi^+$ &$\alpha\vert
10\rangle + \beta\vert 11\rangle + \gamma\vert 00\rangle + \delta
\vert 01\rangle$ &  $U_2\otimes U_0 $
\\
 1  & 0 & $+$ & $-$ & $\psi^-$ &$\alpha\vert
11\rangle + \beta\vert 10\rangle + \gamma\vert 01\rangle + \delta
\vert 00\rangle$ &  $U_2\otimes U_2 $
\\\hline \hline
\end{tabular}\label{table11}
\end{table}
\end{center}
\end{widetext}

Similar to the controlled teleportation \cite{dengpra}, the
original state $\vert \chi\rangle_{ab}$  is an arbitrary one for
two-particle quantum system in the Hilbert space $H^2 \otimes H^2$,
 i.e.,  $\vert \chi\rangle_{ab}=\alpha \vert%
00\rangle_{ab} + \beta \vert 01\rangle_{ab} + \gamma \vert
10\rangle_{ab} + \delta \vert 11\rangle_{ab}$. Moreover, this QSTS
scheme is symmetric as each of the agents can act as the receiver
with the help of the other. In essence, any QSTS scheme can be
used for the controlled teleportation
\cite{dengpra,CTele1,YanFLteleportation2,CTele2} by means that one
of the two  agents acts as the controller and the other recovers
the unknown state according to the information published by the
sender and the controller. That is, this QSTS scheme can be used
to complete the task of controlled teleportation of an arbitrary
two-qubit state more efficient than that in Ref. \cite{dengpra} as
the quantum resource is only two-photon entanglements, not GHZ
states which are not easy for producing
\cite{Mentanglement1,Mentanglement2,Mentanglement3}. On the other
hand, the users should used at least four EPR pairs for sharing
the state in this QSTS scheme. Two EPR pairs ($\vert
\psi^-\rangle_{34}$ and $\vert \psi^-\rangle_{56}$) are used to
transfer the original state and the other two pairs ($\vert
\psi^-\rangle_{12}$ and $\vert \psi^-\rangle_{78}$) are used to
set up the quantum channel between the two agents, Bob and
Charlie, with the control of the sender Alice.

\section{circular QSTS scheme with entanglement swapping}
\label{section3}
In the QSTS scheme discussed above, the sender
Alice should provide the resource for Bob and Charlie to set up
the quantum channel, which will cost Alice a lot of quantum
resource when the number of the agents increases largely. If the
topological structure of the QSTS is circular, the resource can be
reduced greatly. In this way, Alice shares the two-photon
entanglement $\vert \psi^-\rangle_{34}$ with Bob, and $\vert
\psi^-\rangle_{56}$ with Charlie, and then Bob shares the
entanglement $\vert \psi^-\rangle_{78}$ with Charlie, shown in
Fig.2. After the measurements on the photons a and 3, and b and 5,
the original state $\vert \chi\rangle_{ab}$ is transferred to the
photons 4 and 6. That is, the quantum information, the unknown
state, is split by Bob and Charlie. The entanglement $\vert
\psi^-\rangle_{78}$ is just used to transfer the unknown state
$\vert \chi\rangle_{ab}$ to one of the two agents with the help of
the other. The entanglement $\vert\psi^-\rangle_{12}$ is not
necessary in this circular QSTS with two agents.

\begin{figure}[!h]
\begin{center}
\includegraphics[width=8cm,angle=0]{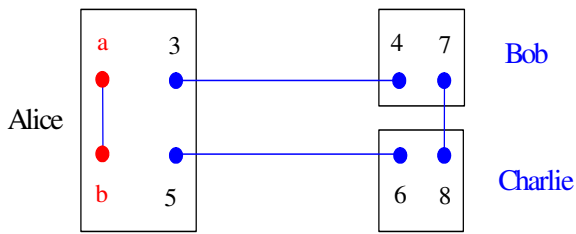} \label{f2}
\caption{ The circular QSTS scheme with entanglement swapping in
the case with two agents. }
\end{center}
\end{figure}

It is straightforwardly to generalize this circular QSTS scheme to
the case with $N$ agents, say Bob$_i$ ($i=1,2,...,N-1$) and Charlie.
As the symmetry, we still assume that Charlie is the agent who will
reconstruct the unknown state with the help of the other $N-1$
agents, Bob$_i$. To this end, Alice should share an EPR pairs $\vert
\psi^- \rangle_{34}$ with Bob$_1$ and another pair $\vert
\psi^-\rangle_{56}$ with Charlie. The $i$-th agent Bob$_i$ shares an
EPR pair $\vert\psi^-\rangle_{2i+5,2i+6}$ with the $(i+1)$-th agent
Bob$_{i+1}$. ($i=1,2,\ldots,N-2$)  [The $(N-1)$-th agent Bob$_{N-1}$
shares the entanglement $\vert \psi^-\rangle_{2N+3,2N+4}$ with
Charlie]. If an agent wants to act as a controller, he performs a
Bell-state measurement on his two photons. That is, all the Bobs
measure their photons and then tell the information of the outcomes
to Charlie for reconstructing the original state $\vert
\chi\rangle_{ab}$ when they cooperate.

Table \ref{table2} gives us the relation between the results and
the local unitary operations. All the notations in Table II are as
same as those in Table \ref{table1}, see Sec.II. We do not exploit
the notation for the EPR pair $\vert\psi^-\rangle_{12}$, then the
total value $V_{total}$ is the sum of the values of the outcomes
obtained by Alice and the $N-1$ controllers Bob$_i$. So does the
total parity $P_{total}$.

\begin{center}
\begin{table}[!h]
\caption{The relation between the local unitary operations and the
results $R_{a3}$, $R_{b5}$, and $R_{2i+5,2i+6}$ ($1\leq i\leq
N-1$). $\Phi_{2N+4,6}$ is the state of the two particles hold in
the hand of Charlie after all the measurements are done by Alice
and Bob$i$; $U_i\otimes U_j$ are the local unitary operations with
which Charlie can reconstruct the unknown state
$\vert\chi\rangle_{ab}$. }
\begin{tabular}{ccccccc}\hline\hline
$V_{total}$  &  $V_{b5}$ &  $P_{b5}$ &  $P_{total}$ &
$\Phi_{2N+4,6}$ &  $U_i\otimes U_j$
\\\hline 
 1  & 1 &  $-$ &  $-$ &   $\alpha\vert
00\rangle + \beta\vert 01\rangle + \gamma\vert 10\rangle + \delta
\vert 11\rangle$ &  $U_0\otimes U_0 $
\\ 
 1  &  1 &  $+$ &  $+$ &
$\alpha\vert 00\rangle - \beta\vert 01\rangle + \gamma \vert
10\rangle - \delta \vert 11\rangle$ &  $U_0\otimes U_1 $
\\ 
0  & 0 & $-$ & $-$ &   $\alpha\vert 00\rangle + \beta\vert
01\rangle - \gamma\vert 10\rangle - \delta \vert 11\rangle$ &
$U_1\otimes U_0 $
\\ 
 0  & 0 & $+$ & $+$ &   $\alpha\vert
00\rangle - \beta\vert 01\rangle - \gamma\vert 10\rangle + \delta
\vert 11\rangle$ &  $U_1\otimes U_1 $
\\ 
 1  & 1 & $-$ & $+$ &   $\alpha\vert
01\rangle + \beta\vert 00\rangle + \gamma  \vert 11\rangle +
\delta\vert 10\rangle$ &  $U_0\otimes U_2 $
\\ 
 1  & 1 & $+$ & $-$ &  $\alpha\vert
01\rangle - \beta\vert 00\rangle + \gamma\vert 11\rangle - \delta
\vert 10\rangle$ &  $U_0\otimes U_3 $
\\ 
 0  & 0 & $-$ & $+$ &  $\alpha\vert
01\rangle + \beta\vert 00\rangle - \gamma\vert 11\rangle - \delta
\vert 10\rangle$ &  $U_1\otimes U_2 $
\\ 
 0  & 0 & $+$ & $-$ &  $\alpha\vert
01\rangle - \beta\vert 00\rangle - \gamma\vert 11\rangle + \delta
\vert 10\rangle$ &  $U_1\otimes U_3 $
\\ 
 0  & 1 & $-$ & $-$ &  $\alpha\vert
10\rangle + \beta\vert 11\rangle + \gamma\vert 00\rangle + \delta
\vert 01\rangle$ &  $U_2\otimes U_0 $
\\ 
 0  & 1 & $+$ & $+$ &  $\alpha\vert
10\rangle - \beta\vert 11\rangle + \gamma \vert 00\rangle - \delta
\vert 01\rangle$ &  $U_2\otimes U_1 $
\\ 
 1  & 0 & $-$ & $-$ &  $\alpha\vert
10\rangle + \beta\vert 11\rangle - \gamma\vert 00\rangle - \delta
\vert 01\rangle$ &  $U_3\otimes U_0$
\\ 
 1  & 0 & $+$ & $+$ &   $\alpha\vert
10\rangle - \beta\vert 11\rangle - \gamma\vert 00\rangle + \delta
\vert 01\rangle$ &  $U_3\otimes U_1 $
\\ 
 0  & 1 & $-$ & $+$ &  $\alpha\vert
11\rangle + \beta\vert 10\rangle + \gamma\vert 01\rangle + \delta
\vert 00\rangle$ &  $U_2\otimes U_2 $
\\ 
 0  & 1 & $+$ & $-$ &   $\alpha\vert
11\rangle - \beta\vert 10\rangle + \gamma\vert 01\rangle - \delta
\vert 00\rangle$ &  $U_2\otimes U_3 $
\\ 
 1  & 0 & $-$ & $+$ &   $\alpha\vert
11\rangle + \beta\vert 10\rangle - \gamma\vert 01\rangle - \delta
\vert 00\rangle$ &  $U_3\otimes U_2 $
\\ 
1  & 0 & $+$ & $-$ &   $\alpha\vert 11\rangle - \beta\vert
10\rangle -\gamma\vert 01\rangle + \delta \vert 00\rangle$ &
$U_3\otimes U_3 $
\\\hline \hline
\end{tabular}\label{table2}
\end{table}
\end{center}

In this circular QSTS scheme, each user should share an EPR pair
with his neighboring one, and he performs Bell-state measurements
on his photons if he wants to act as a controller. Alice's
measurements will transfer the original two-qubit state $\vert
\chi\rangle_{ab}$ to other photons with entanglement swapping. The
measurements done by the  controllers help to set up the quantum
channel for sharing the state with their control. For the view of
producing or measuring a $m$-particle entanglement, this circular
QSTS scheme is an optimal one as it just exploits two-photon
entanglement resource and Bell-state measurements. As almost all
of the photons are useful for the quantum communication, its
efficiency for qubits approaches the maximal value. Same as the
QSTS scheme discussed in section \ref{section2}, any one of the
agents cannot obtain the quantum information, the unknown
two-qubit state unless he cooperate with all of the other agents
even though Alice published the results of her measurements.

\section{Discussion and summary}
\label{section4} As discussed in Refs. \cite{HBB99,KKI}, if Alice
can prevent the dishonest man (no more than one) in the agents
from eavesdropping the quantum secret, the process for sharing an
unknown state is secure for any eavesdropper. In these two QSTS
schemes, their security depends on the process for setting  up the
quantum channel (sharing the maximally entangled states), i.e.,
the EPR pairs. Certainly, it is difficult for two users to share
an EPR pairs securely, but easy to share a sequence of EPR pairs
\cite{twostep,guoatom}. In a noise channel, the parties can
exploit entanglement purification \cite{qpap1,qpap2} to distill
some maximally entangled states for improving the security of
quantum communication. In this way, these two QSTS schemes for
sharing an arbitrary two-qubit states are secure. Another feature
of these two QSTS schemes is that two-particle Bell-state
measurements are required, which is more efficient than those with
$m$-particle joint measurement ($m>2$).

In summary, we present two QSTS schemes for sharing an arbitrary
two-qubit state $\vert \chi\rangle_{ab}=\alpha \vert
00\rangle_{ab} + \beta \vert 01\rangle_{ab} + \gamma \vert
10\rangle_{ab} + \delta \vert 11\rangle_{ab}$ based on
entanglement swapping with EPR pairs and Bell-state measurements.
One is based on the quantum channel with four EPR pairs shared in
advance, the other is based on a circular topological structure.
Any one in the agents has the choice to reconstruct the original
state $\vert \chi\rangle_{ab}$ with the help of the others'.
Moreover the circular QSTS scheme reduces the quantum resource
needed largely when the number of the agents is large. Almost all
the EPR pairs can be used for quantum communication in those two
schemes, their efficiency for qubits approaches the maximal value,
same as Refs. \cite{Peng,dengmQSTS,dengpra}. They are more
convenient in application than the other schemes existing as they
require only two-qubit entanglements and two-qubit joint
measurements for sharing an arbitrary two-qubit state.

\bigskip
\section*{ACKNOWLEDGEMENTS}
This work is supported by the National Natural Science Foundation
of China under Grant Nos. 10447106, 10435020, 10254002, A0325401
and 10374010, and Beijing Education Committee under Grant No.
XK100270454.

\end{document}